\documentclass[sigconf]{acmart}

\usepackage{pifont}
\usepackage{booktabs} % For formal tables
\usepackage{amsmath}
\usepackage{multirow}
\usepackage{flushend}
\usepackage{enumitem}
\usepackage[T1]{fontenc}
\usepackage[utf8]{inputenc}
\usepackage{algorithmic}
\usepackage[linesnumbered,ruled,vlined]{algorithm2e}
\usepackage[switch]{lineno}
\usepackage{eqparbox}
\usepackage[nopar]{lipsum}
\usepackage{makecell}

\usepackage{xcolor}

\DeclareMathOperator*{\argmin}{arg\,min}

\newtheorem*{proof*}{Proof}
\newtheorem*{claim*}{}

\newcolumntype{C}[1]{>{\centering\let\newline\\\arraybackslash\hspace{0pt}}m{#1}}
\newcommand\ChangeRT[1]{\noalign{\hrule height #1}}

\copyrightyear{2022}
\acmYear{2022}
\setcopyright{acmlicensed}\acmConference[SIGIR '22]{Proceedings of the 45th
International ACM SIGIR Conference on Research and Development in
Information Retrieval}{July 11--15, 2022}{Madrid, Spain}
\acmBooktitle{Proceedings of the 45th International ACM SIGIR Conference on
Research and Development in Information Retrieval (SIGIR '22), July 11--15,
2022, Madrid, Spain}
\acmPrice{15.00}
\acmDOI{10.1145/3477495.3531799}
\acmISBN{978-1-4503-8732-3/22/07}

\begin{document}
\fancyhead{}

\title{Distill-VQ: Learning Retrieval Oriented Vector Quantization By Distilling Knowledge from Dense Embeddings}

\author{
%   \alignauthor 
    Shitao Xiao$^{\textbf{{\tiny\ding{171}}}}$, Zheng Liu$^{\textbf{{\tiny\ding{168}}}}$, Weihao Han$^{\textbf{{\tiny\ding{169}}}}$, Jianjin Zhang$^{\textbf{{\tiny\ding{169}}}}$,
    Defu Lian$^{\textbf{{\tiny\ding{170}}}}$,  
    Yeyun Gong$^{\textbf{{\tiny\ding{168}}}}$,
    Qi Chen$^{\textbf{{\tiny\ding{168}}}}$,
    Fan Yang$^{\textbf{{\tiny\ding{168}}}}$,
    Hao Sun$^{\textbf{{\tiny\ding{169}}}}$,
    Yingxia Shao$^{\textbf{{\tiny\ding{171}}}}$,
    Xing Xie$^{\textbf{{\tiny\ding{168}}}}$
}

\affiliation{\ding{171}: 
\institution{Beijing University of Posts and Telecommunications \country{China}}} 
\affiliation{\ding{168}: 
\institution{Microsoft Research Asia \country{China}}} 
\affiliation{\ding{169}: 
\institution{Microsoft Search Technology Center Asia \country{China}}} 
\affiliation{\ding{170}: 
\institution{University of Science and Technology of China \country{China}}} 

\email{{zhengliu,weihan,jianjzh,yegong,cheqi,fanyang,hasun,xingx}@microsoft.com}
\email{{stxiao,shaoyx}@bupt.edu.cn}
\email{liandefu@ustc.edu.cn}

\renewcommand{\shortauthors}{Xiao and Liu, et al.}
\renewcommand{\authors}{Shitao Xiao, 
Zheng Liu, Weihao Han, Jianjin Zhang, Defu Lian,  Yeyun Gong, Qi Chen, Fan Yang, Hao Sun, Yingxia Shao, Xing Xie}

\begin{abstract}
  Vector quantization (VQ) based ANN indexes, such as Inverted File System (IVF) and Product Quantization (PQ), have been widely applied to embedding based document retrieval thanks to the competitive time and memory efficiency. Originally, VQ is learned to minimize the reconstruction loss, i.e., the distortions between the original dense embeddings and the reconstructed embeddings after quantization. Unfortunately, such an objective is inconsistent with the goal of selecting ground-truth documents for the input query, which may cause severe loss of retrieval quality. Recent works identify such a defect, and propose to minimize the retrieval loss through contrastive learning. {However, these methods intensively rely on queries with ground-truth documents, whose performance is limited by the insufficiency of labeled data.} 
  
  In this paper, we propose Distill-VQ, which unifies the learning of IVF and PQ within a knowledge distillation framework. In Distill-VQ, the dense embeddings are leveraged as ``teachers'', which predict the query's relevance to the sampled documents. The VQ modules are treated as the ``students'', which are learned to reproduce the predicted relevance, such that the reconstructed embeddings may fully preserve the retrieval result of the dense embeddings. By doing so, Distill-VQ is able to derive substantial training signals from the massive unlabeled data, which significantly contributes to the retrieval quality. We perform comprehensive explorations for the optimal conduct of knowledge distillation, which may provide useful insights for the learning of VQ based ANN index. We also experimentally show that the labeled data is no longer a necessity for high-quality vector quantization, which indicates Distill-VQ's strong applicability in practice. The evaluations are performed on MS MARCO and Natural Questions benchmarks, where Distill-VQ notably outperforms the SOTA VQ methods in Recall and MRR. 
  Our code is avaliable at https://github.com/staoxiao/LibVQ.
%   Our source code, data, and models will all be open to public.
  
%   In this paper, we propose Distill-VQ where the vector quantization is learned by distilling knowledge from the dense embeddings. Particularly, the dense embeddings are utilized as ``teachers'', which predict the query's relevance to a group of sampled documents. The VQ module is learned to reproduce the predicted relevance, such that the reconstructed embeddings may fully preserve the retrieval result of the dense embeddings. Compared with the existing methods, Distill-VQ may derive substantial training signals from the massive unlabeled data, which gives rise to notable improvement of the retrieval quality. It is also interesting to find that Distill-VQ can be effectively learned without using any labeled data, which indicates its strong applicability in practice. Comprehensive experimental studies are performed on two representative document retrieval benchmarks: MS MARCO and Natural Questions, where Distill-VQ outperforms the SOTA VQ approaches with notable advantages. Our source code, data, and models will all be open to public upon acceptance of the paper. 
\end{abstract}

\vspace{-5pt}


\vspace{-5pt}
% \keywords{Vector Quantization, Knowledge Distillation}

\maketitle

\vspace{-8pt}
\section{Introduction}

\begin{figure*}[t]
\centering
\includegraphics[width=0.98\textwidth]{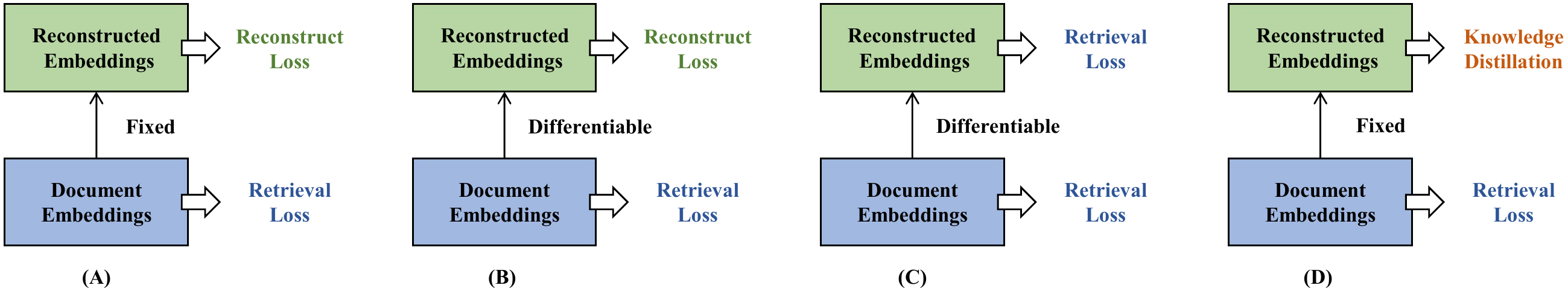}
\vspace{-10pt}
\caption{\small Development of vector quantization. (A): the reconstruction loss is minimized for the fixed document embeddings and reconstructed embeddings; e.g., PQ \cite{jegou2010product}, OPQ \cite{ge2013optimized}. (B): reconstructed embeddings are end-to-end differentiable and learned for reconstruction loss minimization; e.g., DPQ \cite{chen2020differentiable}. (C): reconstructed embeddings are end-to-end differentiable and learned for retrieval loss minimization; e.g., MoPQ \cite{xiao2021matching}, JPQ \cite{zhan2021jointly}, RepCONC \cite{zhan2021learning}. (D): quantized embeddings are learned by knowledge distillation based on fixed document embeddings (Distill-VQ).} 
\vspace{-10pt}
\label{fig:1}
\end{figure*}

The embedding based retrieval (EBR) of documents plays an important role in many web applications, such as search engines and recommender systems \cite{gillick2019learning,karpukhin2020dense,chang2020pre,huang2020embedding}: given an input query, the relevant documents are selected from the entire corpus based on embedding similarity. Knowing that brute-force linear scan is temporally infeasible, the embeddings need to be organized by ANN index in order to support real-world information retrieval. For the past decade, the vector quantization (VQ) techniques, e.g., Inverted File System (IVF) \cite{jegou2011searching,babenko2014inverted,baranchuk2018revisiting} and Product Quantization (PQ) \cite{jegou2010product,subramanya2019diskann,xiao2021matching}, are widely applied, which enables ANN to be performed with competitive time and memory efficiency. 

$\bullet$ \textbf{Typical learning paradigms of VQ}. VQ is typically learned to minimize the reconstruction loss, i.e., the minimization of distortions (e.g., $l$2 distance) between the original dense embeddings and the reconstructed embeddings from quantization. Unfortunately, such an operation is inconsistent with the goal of retrieving ground-truth documents for the query, which may cause severe loss of retrieval performance. In recent works \cite{xiao2021matching,zhan2021jointly}, the dense embeddings and VQ are jointly learned to minimize the post-quantization retrieval loss. However, the existing methods typically rely on contrastive learning, which calls for a tremendous amount of queries labeled with ground-truth documents. As a result, it may suffer from inferior performance due to the insufficiency of labeled data.  

% However, the jointly learning methods suffer from low data-effectiveness, which are prone to inferior recall when labeled data is insufficient. 

$\bullet$ \textbf{Our Solution}. In this paper, we propose \textbf{Distill-VQ}, which jointly learns IVF and PQ via knowledge distillation. In Distill-VQ, we leverage well-trained dense embeddings\footnote{We target on optimizing the retrieval performance resulted from vector quantization. How to train dense embeddings is beyond the discussion of this paper.} as the teachers: for each query, the teachers will predict its relevance scores towards a set of candidate documents. The predicted scores are used as the indicators of how likely the corresponding documents are the ground-truth to the query. The quantization modules are treated as the students, which are learned to reproduce the teachers' predictions, such that the dense embeddings' retrieval performance can be effectively preserved. Compare with the existing methods, Distill-VQ gives rise to the following merits. 

% \nocite{*}

First and foremost, Distill-VQ is able to exploit the data more effectively, which will significantly contribute to the retrieval quality. The current joint learning methods are typically based on contrastive learning and driven by labeled data: the model is learned to identify the document labeled as ground-truth to each query, where the remaining documents within the corpus are equally treated as negative samples. In contrast, Distill-VQ is to learn from the teachers (i.e., the well-trained dense embeddings), thus relaxing the limitation from labeled data. During the knowledge distillation process, the high-relevance documents to each query, whether labeled as ground-truth or not, will probably be given high relevance scores from the teachers. As a result, such documents will virtually serve as positive samples, which are huge supplements to the original labeled dataset. Besides, instead of treating all remaining documents equally as negative samples, Distill-VQ will penalize more for the documents which are predicted to be less relevant with the query. Therefore, it may also alleviate the false negative problems existed in the conventional methods. 

In addition, Distill-VQ also contributes to {higher applicability} while working with those off-the-shelf document embeddings, e.g., CDSSM \cite{kim-2014-convolutional} and SBERT \cite{reimers2019sentence}. As discussed, the existing joint learning methods typically learn dense embeddings and VQ at the same time leveraging massive-scale of labeled data. However, it remains little explored of how to optimize the VQ's performance on top of well-trained dense embeddings. Distill-VQ naturally alleviates this problem as it is designed to work with fixed document embeddings. Besides, knowing that the model is learned from knowledge distillation, the labeled data is no longer a necessity. In fact, it is experimentally verified that Distill-VQ may generate high-quality vector quantization without using ground-truth documents, which significantly relaxes the requirement on training data. 

{We perform comprehensive explorations for the optimal conduct of knowledge distillation, including the document sampling strategies (i.e., which documents to distill the teachers' knowledge), and the function to preserve teachers and students' similarity (which defines how to distill teachers' knowledge from their predictions to the sampled documents). It is found that distilling knowledge from a sufficient large number of Top-K documents (the Top-K relevant documents to each query predicted by the teachers) with functions enforcing ranking order invariance, e.g., ListNet \cite{cao2007learning}, may produce the most effective training outcome. Our evaluations are performed on two large-scale benchmarks: MS MARCO and Natural Questions, where Distill-VQ outperforms the SOTA vector quantization methods by notable margins. We also integrate Distill-VQ's output into FAISS \cite{johnson2019billion}, which significantly improves the performance of its original VQ-based indexes, e.g., IVFPQ, IVF-Flat; and enables our method to be directly applicable to real-world applications. To summarize, our work is highlighted with the following points.}

% Our experimental studies are performed with two large-scale document retrieval benchmarks: MS MARCO and Natural Questions, where Distill-VQ outperforms the SOTA vector quantization methods by notable margins. Besides, we also provide comprehensive explorations for the effective conduct of knowledge distillation, including the sampling strategies for candidate documents (i.e., which documents to distill the teachers' knowledge), and the function for the measurement of teachers and students' similarity (i.e., how to distill teachers' knowledge from their predictions towards the candidate documents). We've integrated Distill-VQ with FAISS library \cite{johnson2019billion}, which significantly improves the performance of its original indexes, such as IVFPQ and IVF-Flat, and enables our method to be directly applicable to real-world document retrieval applications. Corresponding implementations will also be released to public in order to facilitate the progress in the community. To summarize, our work is highlighted by the following points. 

\vspace{-8pt}
\setlist[itemize]{leftmargin=5.0mm}
\begin{itemize}
    \item We propose a novel framework Distill-VQ, which jointly learns IVF and PQ for the optimization of their retrieval performance. 
    
    \item By distilling knowledge from the well-trained dense embeddings, Distill-VQ achieves much more effective exploitation of the data, and contributes to a higher applicability when working with off-the-shelf embeddings. 
    
    % Experiments show that high-quality vector quantization can be generated without purely with unlabeled data. 
    
    \item Comprehensive explorations are performed for the optimal conduct of Distill-VQ, including the document sampling strategies and the functions to enforce teachers and students' consistency. 
    
    \item Experimental studies on MS MARCO and Natural Questions benchmarks verify the effectiveness of Distill-VQ. 
\end{itemize}

\begin{figure*}[t]
\centering
\includegraphics[width=0.88\textwidth]{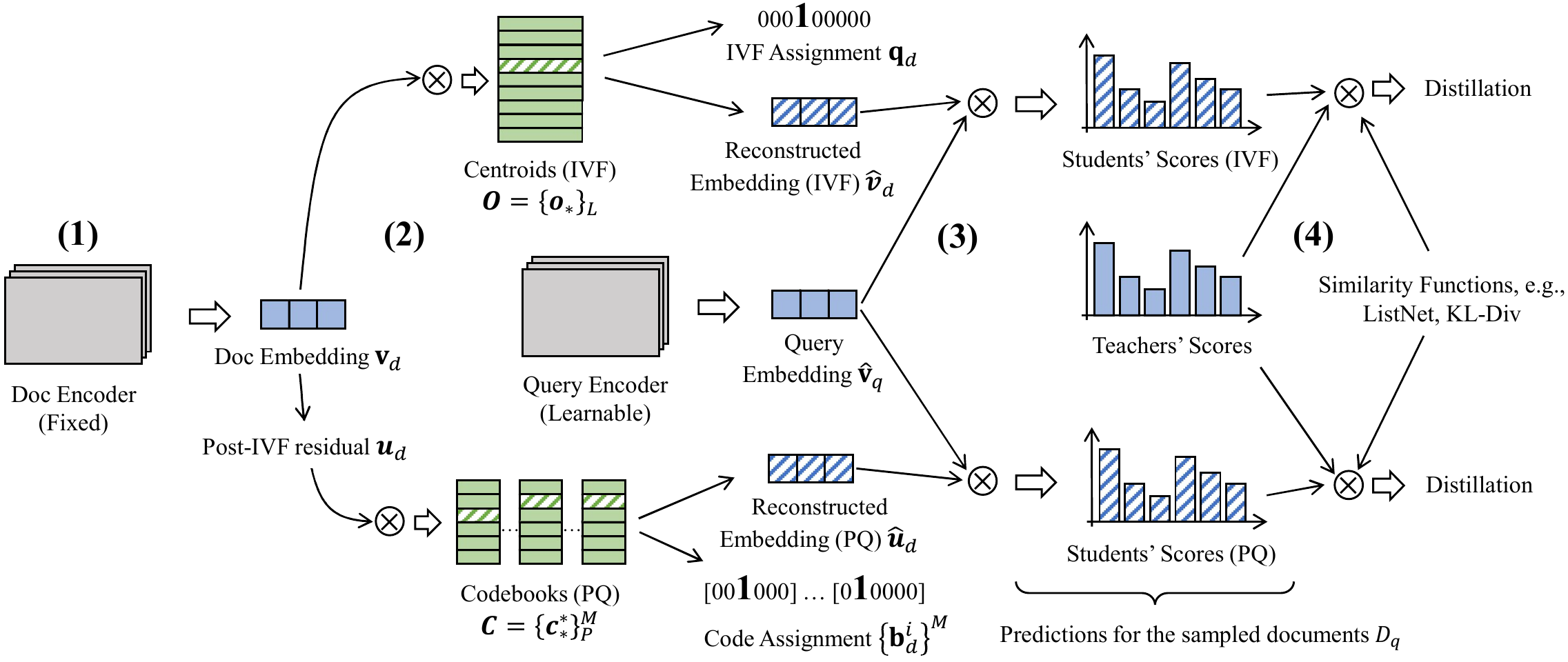}
\vspace{-10pt}
\caption{\small Distill-VQ. (1) The Doc Encoder (well-trained and fixed) infers the embeddings for all documents in the offline stage. (2) Each document embedding is assigned to IVF and PQ (the assigned entries are stripped); then, the embedding is reconstructed w.r.t. the assignments of IVF and PQ, respectively. (3) The Query Encoder infers the query embedding, which will interact (inner-product in our work) with the reconstructed embeddings for the relevance scores with the document. For each query: the relevance scores are computed for all the sampled documents. (4) The knowledge distillation is performed w.r.t. the relevance scores from teachers (computed in the offline stage).} 
\vspace{-10pt} 
\label{fig:2}
\end{figure*}

\vspace{-5pt}
\section{Related Works}\label{sec:ref}

%% EBR: model training
%% ANN: graph and VQ
%% VQ: IVF, PQ, OPQ, DPQ ...
%% Objective: reconstruct, retrieval

Embedding based retrieval has been widely applied in practice, such as search engines \cite{shen2014learning,liu2021pre}, question answering \cite{multi_grained_query,qu2021rocketqa, QA_DSE}, online advertising \cite{lu2020twinbert,li2021adsgnn}, and content-based recommender systems \cite{xiao2021training,li2021embedding}. Knowing that the documents need to be retrieved from a large-scale corpus, where brute-force linear scan will be temporally infeasible, it calls for approximate nearest neighbour search (ANN) \cite{li2019approximate} such that documents with high embedding similarities can be efficiently selected. The ANN index needs to trade-off three key factors: time cost, memory usage, and recall rate. That's to say, it is expected to recall high-quality candidates with the minimum consumption of time and memory. In recent years, the VQ based ANN indexes (e.g., IVFADC \cite{jegou2011searching}, IMI \cite{babenko2014inverted}, IVFADC-G-P \cite{baranchuk2018revisiting}, DiskANN \cite{subramanya2019diskann}, SPANN \cite{chen2021spann}) have been widely recognized as effective solutions to realize this objective. For the family of VQ based ANN indexes: the original dense embeddings are quantized based on the codec of index, including the IVF's centroids, and the PQ's codebooks. With this operation, the document can be expressed as binary codes, which is much lighter than the original dense embeddings. What's more, the retrieval process can be significantly accelerated based on the inverted file system and the approximate distance computation from the codebooks.

%% OPQ, DPQ, DRN, Poeem
%% Prove wrong by MoPQ

For the past decade, the learning of VQ has passed through different stages, with training objective evolving from reconstruction loss minimization to retrieval loss minimization. The detailed development is discussed as follows. 

$\bullet$ \textbf{Reconstruction loss minimization}. VQ is originally learned to minimize the reconstruction loss, i.e., the minimization of the distortions (e.g., the $l$2 distance) between the reconstructed embeddings and the original dense embeddings. Typical methods work with fixed document embeddings, and rely on unsupervised approach, such as clustering (e.g., $k$-means) or hierarchical clustering, for the learning of codebooks \cite{jegou2010product,jegou2011searching} and inverted indexes \cite{jegou2011searching,baranchuk2018revisiting,chen2021spann}, as Figure \ref{fig:1} (A). Knowing that the distribution of the original dense embeddings can be unfavorable for quantization, people propose OPQ \cite{ge2013optimized}, where the optimal rotation matrix is learned for a more effective reduction of the reconstruction loss. Later on, it is also proposed to jointly learn vector quantization and embedding models in an end-to-end differentiable way \cite{yue2016deep,cao2017deep,chen2020differentiable,zhang2021joint} as Figure \ref{fig:1} (B), where the reconstruction loss can be further reduced.  

%% Retrieval oriented: MoPQ, JPQ, RepCONC
$\bullet$ \textbf{Retrieval loss minimization}. One potential problem about the above works is that VQ is learned purely based on document embeddings, whereas the relationships with queries are not taken into consideration. Besides, it is also analyzed in \cite{xiao2021matching} that the reduction of the reconstruction loss does not necessarily improve the retrieval quality. Realizing such defects, the latest works (e.g., MoPQ \cite{xiao2021matching}, JPQ \cite{zhan2021jointly}, RepCONC \cite{zhan2021learning}) propose to jointly learn VQ and the embedding models via contrastive learning, in which the discrimination of the ground-truth documents for each query can be optimized (referred as retrieval loss minimization), as Figure \ref{fig:1} (C). Compared with the previous minimization of the reconstruction loss, the retrieval loss minimization achieves substantial improvements in terms of the retrieval performance. 

$\bullet$ \textbf{Distinctions of our work}. Distill-VQ is similar with the existing works as it also targets on optimizing the retrieval quality rather than simply minimizing the reconstruction loss. However, Distill-VQ is differentiated in the following aspects, as Figure \ref{fig:1} (D). First and foremost, Distill-VQ is learned from well-trained dense embeddings via knowledge distillation, while the existing methods rely on contrastive learning. Secondly, Distill-VQ relaxes the requirement on labeled data, and can be effectively trained purely with unlabeled data. Thirdly, Distill-VQ works with fixed document embeddings instead of end-to-end optimization of the whole model.

%% Knowledge distillation: KD, noisy student, TAS-balance, RocketQAv2, AR-G
%% Why KG works: augmentation of labels, false negatives when using ance hard negatives 

$\bullet$ \textbf{Knowledge Distillation}. Finally, it should be noted that knowledge distillation (KD) \cite{hinton2015distilling} plays the fundamental role in Distill-VQ. KD used to be a powerful tool for model compression \cite{sun2019patient,sanh2019distilbert,jiao2019tinybert}, where the student model (lightweight) is learned to imitate the predictions from the teacher model (heavyweight). In recent years, it becomes increasingly popular to apply KD for representation learning. For example, in \cite{hofstatter2021efficiently,ren2021rocketqav2,zhang2021adversarial}, the representation models are distilled from the relevance scores predicted by the re-ranking models; and in \cite{xie2020self,chen2020big}, the representation models are pretrained by distilling knowledge from the pseudo labels annotated by the teacher models. It is generally believed that KD contributes to the representation learning thanks to the exploitation of massive unlabeled data \cite{xie2020self} and label smoothing \cite{yuan2020revisiting}. Distill-VQ inherits such merits of KD, thus becoming more effective in preserving the dense embeddings' retrieval performance.

\vspace{-8pt}
\section{Distill-VQ}
%% Overview (IVFPQ): encoding, quantization, ADC; generic: IVFPQ -> IVF and PQ
%% Quantization and Reconstruction: - IVF / - PQ
%% Distillation: teacher score, fixed assignment, update query encoder & index
%% Distillation: candidate distribution, distillation function

The framework of Distill-VQ is shown as Figure \ref{fig:2}, where both IVF and PQ modules are optimized. As a result, the developed techniques can be directly applied to generic VQ-based ANN indexes, such as PQ, IVFPQ, IVF-Flat, IVFADC-G-P, DiskANN, SPANN, etc. The workflow of Distill-VQ is divided into four steps. Firstly, Distill-VQ works with a well-trained document encoder, which generates the embeddings for all documents in the offline stage. The document embeddings are fixed in Distill-VQ, which longer take any computation cost in quantization learning. Secondly, each document embedding is assigned to its nearest centroid in IVF and the nearest codeword in each codebook of PQ. The assignment results are represented in binarized forms (i.e., one-hot vectors indicating the assigned ID), and used to generate {the reconstructed embeddings} for the document. Thirdly, the query encoder is used to generate query embedding, which will interact (inner-product) with the reconstructed document embedding for the relevance score. Finally, the knowledge distillation is performed, which enables the relevance scores computed by students (i.e., the reconstructed embeddings) to imitate the predictions from the teachers (i.e., the well-trained dense embeddings of the queries and documents).

\vspace{-8pt}
\subsection{Quantization and Reconstruction} 
The document embeddings are quantized and reconstructed via the following steps. Firstly, the document embedding $\mathbf{v}_d$ is generated for each document $d$ from the well-trained document encoder. Secondly, the IVF's centroids  ($\mathbf{O} = \{\mathbf{o}_i \in \mathbb{R}^h \}_{L}$; $L$: size of IVF) and PQ's codebooks ($\mathbf{C} = \{\mathbf{c}_i^j \in \mathbb{R}^{h/M} \}_{P}^{M}$; $P$: number of codewords per codebook; $M$: number of codebooks) are initialized from the document embeddings based on vanilla IVFPQ \cite{jegou2011searching}:
\begin{equation}\label{eq:1}
    \mathbf{O}, \mathbf{C} \leftarrow \mathrm{IVFPQ}(\{\mathbf{v}_d\}_D). 
\end{equation} 
The document embeddings can be quantized and reconstructed on top of the initialized IVF and PQ. For IVF, each document embedding is quantized by assigning to the closest centroid (using $l$2 distance); and gets reconstructed as $\mathbf{\hat{v}}_d$ based on the assigned centroid\footnote{{Considering the high time cost of reassignment over large-scale IVF centroids, we fix the assignment of IVF centroids in the training process.}}:
\begin{equation}\label{eq:2}
        \mathbf{q}_d \leftarrow \argmin_{i} \| \mathbf{v}_d - \mathbf{o}_i \|; ~
    \mathbf{\hat{v}}_d \leftarrow \mathbf{o}_{\mathbf{q}_d}.
\end{equation}
For PQ, the post-IVF residue $\mathbf{u}_d$ is computed for each document embedding based on its IVF assignment: $\mathbf{u}_d \leftarrow \mathbf{v}_d - \mathbf{\hat{v}}_d$. Then, the post-IVF residue is quantized into $M$ one-hot vectors $\{\mathbf{b}_d^i\}^M$ by assigning each segment of $\mathbf{u}_d^{i} \in \{\mathbf{u}_d^{i}\}^{1...M}$ to its closest codeword in the corresponding codebook, and gets reconstructed as $\mathbf{\hat{u}}_d$ by concatenating all the assigned codewords: 
\begin{equation}\label{eq:3}
    \mathbf{b}_d^i \leftarrow \argmin_{j} \| \mathbf{u}^i_d - \mathbf{c}_j^i \|; ~
    \mathbf{\hat{u}}_d = \mathrm{concat}([\mathbf{c}_{\mathbf{b}_d^1}^1, ... , \mathbf{c}_{\mathbf{b}_d^M}^M]). 
\end{equation} 
% Note that the IVF assignment $\mathbf{q}_d$ will be fixed in the subsequent knowledge distillation process (whereas the codebook assignment $\mathbf{b}^i_d$ is dynamically determined). In other words, $\mathbf{\hat{v}}_d$ will always be corresponding to the same centroid

% . However, the reconstructed embeddings ($\mathbf{\hat{u}}_d$, $\mathbf{\hat{v}}_d$) will be trained for the optimal retrieval performance, where the centroids and codebooks are the learnable parameters. 

\subsection{Knowledge Distillation}
%% Distillation formulation. Input: teacher, student, query, documents. Query embedding. Independent distill. Training cost analysis. 
%% Distillation functions
%% Distillation distributions 

The reconstructed embeddings are learned from knowledge distillation, where the well-trained embeddings for the document ($\mathbf{v}_d$) and query ($\mathbf{v}_q$) are utilized as the \textbf{Teachers}. Particularly, for each query ($q$), the teachers' scores ($T_q$) are computed for its sampled candidate documents ($d \in D_q$) based on $\mathbf{v}_d$ and $\mathbf{v}_q$:
\begin{equation}\label{eq:4}
    T_q \leftarrow \{ \langle \mathbf{v}_q, \mathbf{v}_d \rangle | d \in D_q \},
\end{equation}
where ``$\langle\cdot\rangle$'' is the inner-product operator. The reconstructed embeddings are treated as the \textbf{Students}, which will generate their scores ($S^{\alpha}_q$ and $S^{\beta}_q$) towards each query's candidate documents: 
\begin{equation}\label{eq:5}
    \begin{split}
        \text{IVF}: ~ S^{\alpha}_q \leftarrow \{ \langle \mathbf{\tilde{v}}_q, \mathbf{\hat{v}}_d \rangle | d \in D_q \}; \\
        \text{PQ}: ~ S^{\beta}_q \leftarrow \{ \langle \mathbf{\tilde{v}}_q, \mathbf{\hat{u}}_d+\mathbf{\hat{v}}_d \rangle | d \in D_q \}. 
    \end{split}
\end{equation}
Note that the query encoder is still learnable in knowledge distillation. Therefore, the query embedding $\mathbf{\tilde{v}}_q$ will become different from its original value $\mathbf{v}_q$ (i.e., the one used for computing teachers' scores). The students' scores are required to preserve the teachers' predictions w.r.t. a certain similarity function ``$f(\cdot)$''. Finally, the objective for the knowledge distillation is formulated as:
\begin{equation}\label{eq:6}
    \min_{\mathbf{O},\mathbf{C},\mathbf{\tilde{v}}_q} - \sum_q \Big( f(T_q, S^{\alpha}_q | D_q) + f(T_q, S^{\beta}_q | D_q)\Big).
\end{equation}
The above optimization problem involves two groups of learnable parameters: the IVFPQ related parameters $\mathbf{O}$ and $\mathbf{C}$, and the query encoder (which generates $\mathbf{\tilde{v}}_q$). Compared with the conventional joint learning methods \cite{xiao2021matching,zhan2021jointly}, the document embedding $\mathbf{v}_d$ is fixed in Distill-VQ, meaning that the document encoder is excluded from the training process. Considering that the document is usually much larger than the query in terms of sequence length, and there are multiple candidate documents sampled for each query, such a difference will lead to a major {saving of the computation cost}. 

\begin{algorithm}[t]
\caption{\small Distill-VQ}\label{alg:1}
    \LinesNumbered 
    \SetKwInOut{KwIn}{Input}
    \SetKwInOut{KwOut}{Output}
    \KwIn{Well-trained dense embeddings: $\{\mathbf{v}_d\}_D$, $\{\mathbf{v}_q\}_Q$} 
    \KwOut{IVF, PQ, updated query encoder}
    \Begin{
        \While{Not converge}{
            \For{$q \in Q$}{
                $D_q \leftarrow$ sample candidate documents from $D$\;
                $T_q \leftarrow$ generate teachers' scores as Eq. \ref{eq:4}\;
                $S^{\alpha}_q, S^{\beta}_q\leftarrow$ generate students' scores as Eq. \ref{eq:5}\;
                Update IVF, PQ and query encoder w.r.t. Eq. \ref{eq:6}\;
            }
        } 
        % Return IVF and PQ\; 
    }
\end{algorithm}

Given the above formulation, the vector quantization is learned through the following iterative process (Alg. \ref{alg:1}): for each query, the candidate documents $D_q$ are sampled from the entire documents $D$; then, teachers' scores are computed based on the well-trained dense embeddings, and the students' scores are computed based on the reconstructed embeddings; finally, the model is updated for the minimization of the distillation loss through gradient descent. 

The knowledge distillation in Alg. \ref{alg:1} involves two critical factors: 1) the \textbf{Similarity Function} $f(\cdot)$, which determines what kind of similarity to be preserved between the teachers and students; and 2) the \textbf{Sampling Strategy} for the candidate documents $D_q$, which identifies the documents where the similarity needs to be preserved. In the next part, we'll make discussions on a series of alternative solutions for both factors, whose impacts will be comprehensively explored in our experimental studies.   

\subsubsection{Similarity Function}\label{sec:sim_fn}
The students may preserve the teachers' predictions in two alternative ways. 
One is to enforce the \textbf{Score Invariance}, where the students are expected to generate the same relevance scores for the candidate documents as predicted by the teachers. It is a quite rigid requirement for the similarity between teachers and students. In this place, we'll introduce the following similarity functions for the enforcement of score invariance. 

$\bullet$ \textbf{Mean Square Error} (\textbf{MSE}). The simplest form of score invariance is to minimize the absolute difference between the scores predicted by the teachers and students:
\begin{equation}\label{eq:7}
    f(T_q, S^*_q | D_q) = {\sum_{d \in D_q} \| \mathbf{s}_{q,d} - \mathbf{t}_{q,d} \|}, 
\end{equation}
where $S^*_q$ are the students' predicted scores, i.e., $S^{\alpha}_q$ or $S^{\beta}_q$; ${\mathbf{s}}_{q,d}$ and ${\mathbf{t}}_{q,d}$ are the students' and teachers' predictions towards each candidate document $d$, as Eq. \ref{eq:4} and \ref{eq:5}. 

$\bullet$ \textbf{Margin MSE}. A modification of MSE is utilized in \cite{hofstatter2021efficiently}: instead of directly minimizing the absolute difference, it minimizes the difference between the margins of scores for two candidate documents $d_1$ and $d_2$, as predicted by the teachers and students:
\begin{equation}\label{eq:8}
    f(T_q, S^*_q | D_q) = \sum_{d_1, d_2 \in D_q} \|
    ({\mathbf{t}}_{q,d_1} - \mathbf{t}_{q,d_2}) - ({\mathbf{s}}_{q,d_1} - \mathbf{s}_{q,d_2}) \|.
\end{equation} 
Margin MSE is a relaxation of MSE, which emphasizes more of the pairwise relationships between the candidate documents rather than the absolute scores. 

Aside from the above functions for the enforcement of score invariance, a more flexible option is to enforce the \textbf{Ranking Order Invariance}, where the students are merely expected to generate the same ranking orders for the candidate documents as predicted by the teachers. It is a highly relaxed requirement for the teachers and students similarity, which will probably become much easier to optimize. In this place, we introduce the following similarity functions to enforce the ranking order invariance. 

$\bullet$ \textbf{RankNet}. The weighted RankNet (RankNet for short) was a pairwise ranking loss proposed in \cite{hofstatter2020improving}. With RankNet, the teachers and students' consistency is to be kept in terms of the pairwise relationship between two candidate documents: 
\begin{equation}\label{eq:12}
    f(T_q, S^*_q | D_q) = - \sum_{d_1, d_2 \in D_q} (\mathbf{t}_{q,d_1} - \mathbf{t}_{q,d_2}) * 
    \log \sigma(\mathbf{s}_{q,d_1} - \mathbf{s}_{q,d_2}). 
\end{equation} 
It can be regarded as a variation of BPR loss \cite{rendle2012bpr} weighted by the teachers' prediction. Aside from the pairwise ranking loss, the functions on listwise similarity may also be leveraged. 

$\bullet$ \textbf{KL-divergence}. The teachers and students' predicted scores towards the candidate documents are normalized into the following distributions (teachers: $\Phi_{q,*}^t$, students: $\Phi_{q,*}^s$):
\begin{equation}\label{eq:9}
\begin{split}
    \Phi_{q,d}^t = \frac{\exp(\mathbf{t}_{q,d})}{\sum_{d \in D_q} \exp(\mathbf{t}_{q,d})}, ~
    \Phi_{q,d}^s = \frac{\exp(\mathbf{s}_{q,d})}{\sum_{d \in D_q} \exp(\mathbf{s}_{q,d})}.
\end{split}
\end{equation}
A sufficient condition for ranking order invariance is that the students may generate the same distribution as the teachers. To this end, the KL-divergence is minimized for the above distributions:
\begin{equation}\label{eq:10}
    f(T_q, S^*_q | D_q) = 
    \text{KL}(\Phi_{q,*}^s || \Phi_{q,*}^t) = -
    \sum_{d \in D_q} \Phi_{q,d}^s \log \frac{\Phi_{q,d}^t}{\Phi_{q,d}^s}.
\end{equation} 

$\bullet$ \textbf{ListNet}. A close variant of KL-divergence is ListNet \cite{cao2007learning}, which is a well-known function for ranking order preservation. In ListNet, the cross-entropy is minimized for the students and teachers' distribution over the candidate documents (as Eq. \ref{eq:9}): 
\begin{equation}\label{eq:11}
    f(T_q, S^*_q | D_q) = \text{H}(\Phi_{q,*}^t, \Phi_{q,*}^s) = - 
    \sum_{d \in D_q} \Phi_{q,d}^t \log {\Phi_{q,d}^s}. 
\end{equation}
Both KL-divergence and ListNet require the listwise similarity between teachers and students. What's more, the teachers and students' similarities on the top-ranked documents can be naturally emphasized by both functions, which enables knowledge distillation to be more consistent with the retrieval metrics.   

\subsubsection{Candidate Document Sampling}\label{sec:doc_sample}
The retrieval performance (reflected by metrics like Recall@K and MRR) is mainly determined by the top-ranked documents, i.e., whether the ground-truth can be covered by the Top-K retrieved documents. Therefore, the students are expected to emphasize more of its similarity with the teachers on such groups of documents. Intuitively, the following sources of candidate documents can be applied for this purpose. 

% To this end, the following sampling sources can be considered. 

$\bullet$ \textbf{Ground-Truth}. One straightforward option is to utilize the {ground-truth} documents of each query, as these documents are the targets to be retrieved. Ideally, the teachers' retrieval performance can be fully preserved if the students may always generate the same rankings for the ground-truth documents (as the teachers). However, it will probably take a huge amount of labeled data to ensure the knowledge distillation quality, which will be a challenging condition in reality. 

$\bullet$ \textbf{Top-K Documents}. Another other option is to select the documents with the Top-K similarities to the query as the candidates (based on the original dense embeddings):
\begin{equation}\label{eq:13}
    D_q = \text{Top-K} \{\langle \mathbf{v}_q, \mathbf{v}_d \rangle|d \in D\}.
\end{equation}
The Top-K documents are plausible for two reasons. Firstly, the students will fully preserve the teachers' retrieval performance as long as they may generate the same Top-K documents. Secondly, it relaxes the limitation from labeled data, since plenty of training samples can be generated for knowledge distillation by getting the Top-K documents of arbitrary queries. 

$\bullet$ \textbf{In-Batch Documents}. Aside from the top-ranked documents, the lower-ranked documents need to be introduced for the contrastive purpose, i.e., to make sure the lower-ranked documents from the teachers are safely excluded from the students retrieval results. In this place, we leverage \textbf{in-batch sampling} \cite{karpukhin2020dense}, where one query will make use of other queries' related documents within the same batch for knowledge distillation. 
% Given that Distill-VQ only needs to encode the queries, the batch size can be 

Finally, we may come up with the following sampling strategies. 

$\bullet$ \textbf{All-Mixed}. The first one still uses labeled data, where the candidate documents are a mixture of all three sources: 1) ground-truth documents, 2) Top-K documents, 3) in-batch documents. 

$\bullet$ \textbf{Top-K and In-Batch}. The second one purely works with the unlabeled data, where the candidate documents are merely a mixture of the Top-K and in-batch documents. 

The impacts of both strategies are extensively analyzed in our experiment, together with explorations of other influential factors, such as the value of K, and how many documents to sample from the Top-K, etc. It is interesting to find that purely working with the unlabeled data, i.e., the Top-K and In-Batch combination, is sufficient to achieve highly competitive knowledge distillation performance. 

% $\bullet$ \textbf{Conclusion}. To summarize, we derive the following conclusion from the experiments: 1) ranking order invariance is more ef

% the similarity functions for ranking order invariance and the candidate documents from the Top-K and In-Batch combinations give rise to more effective knowledge distillation. For our specific settings, the best retrieval perf

% the combination of ListNet and Top-K+In-Batch result in the 

\begin{table}[t]
    \centering
    \small
    % \scriptsize
    % \footnotesize
    \begin{tabular}{p{2.4cm}  C{2.4cm} C{2.4cm} }
    \ChangeRT{1pt} Datasets &
    MS MARCO Passage &  Natural Questions \\
    \hline
    \#Train (original) & 502,939 & 79,168 \\
    \#Train (processed) & 502,939 & 58,880  \\
    \#Dev & 6,980 & 8,757 \\
    \#Test & -- & 3,610 \\
    \#Documents & 8,841,823 & 21,015,324 \\
    \ChangeRT{1pt}
    \end{tabular}
    % \vspace{-5pt}
    \caption{\small Specifications of the datasets.}
    \vspace{-15pt}
    \label{tab:1}
\end{table}

\begin{table*}[h]
    \centering
    \footnotesize
    \begin{tabular}{p{2.0cm} | p{1.6cm} | C{1.2cm} C{1.2cm} C{1.2cm} C{1.2cm} | C{1.2cm} C{1.2cm} C{1.2cm} C{1.2cm} }
    \ChangeRT{1pt}  & & 
    \multicolumn{4}{c|}{\textbf{MS MARCO Passage (Dev)}} & \multicolumn{4}{c}{\textbf{Natural Questions (Test)}}   \\
    \cmidrule(lr){1-1}
    \cmidrule(lr){2-2}
    \cmidrule(lr){3-6}
    \cmidrule(lr){7-10}
     \textbf{Encoder} & \textbf{Method} & 
     \textbf{MRR@10} & \textbf{R@10} & \textbf{R@50} & \textbf{R@100} & \textbf{MRR@10} & \textbf{R@10} & \textbf{R@50} & \textbf{R@100}  \\
     \hline
     \multirow{7}{*}{AR2-G \cite{zhang2021adversarial}} 
     & IVFPQ & 0.2403 & 0.4545 & 0.6615 & 0.7325 & 0.2823 & 0.5002 & 0.6764 & 0.7240 \\
     & IVFOPQ & 0.3448 & 0.6013 & 0.7881 & 0.8385 & 0.5505 & 0.7260 & 0.8047 & 0.8279 \\
     & ScaNN & 0.2670 & 0.4880 & 0.6953 & 0.7963 & 0.4137 & 0.6470 & 0.7742 & 0.8124 \\
     & Poeem & 0.3448 & 0.6077 & 0.7975 & 0.8487 & 0.5588 & 0.7443 & 0.8246 & 0.8445 \\
     & JPQ & 0.3451 & 0.6107 & 0.8028 & 0.8558 & 0.5666 & 0.7512 & 0.8282 & 0.8504 \\
     & RepCONC & 0.3449 & 0.6106 & 0.7983 & 0.8497 & 0.5500 & 0.7382 & 0.8141 & 0.8373 \\
     & MoPQ & 0.3471 & 0.6119 & 0.8037 & 0.8564 & 0.5835 & 0.7634 & 0.8412 & 0.8606 \\
     & Distill-VQ & \textbf{0.3607}$^\dagger$ & \textbf{0.6276}$^\dagger$ & \textbf{0.8221}$^\dagger$ & \textbf{0.8719}$^\dagger$ & \textbf{0.6235}$^\dagger$ & \textbf{0.7927}$^\dagger$ & \textbf{0.8627}$^\dagger$ & \textbf{0.8783}$^\dagger$ \\
     \hline
     \multirow{7}{*}{CoCondenser \cite{gao2021unsupervised}} 
     & IVFPQ & 0.2252 & 0.4380 & 0.6509 & 0.7202 & 0.3871 & 0.6177 & 0.7642 & 0.8047 \\
     & IVFOPQ & 0.3340 & 0.5947 & 0.7810 & 0.8325 & 0.5447 & 0.7362 & 0.8385 & 0.8626 \\
     & ScaNN & 0.2470 & 0.4602 & 0.6715 & 0.7442 & 0.4520 & 0.6540 & 0.7831 & 0.8193 \\ 
     & Poeem & 0.3363 & 0.6026 & 0.7908 & 0.8403 & 0.5522 & 0.7476 & 0.8354 & 0.8573 \\
     & JPQ & 0.3360 & 0.5954 & 0.7810 & 0.8335 & 0.5436 & 0.7448 & 0.8335 & 0.8551 \\
     & RepCONC & 0.3319 & 0.5895 & 0.7798 & 0.8342 & 0.5544 & 0.7495 & 0.8356 & 0.8602 \\
     & MoPQ & 0.3370 & 0.6056 & 0.8019 & 0.8542 & {0.5689} & 0.7549 & 0.8395 & 0.8627 \\
    %  0.3523	0.6212	0.8086	0.8612	0.5781	0.7654	0.8543	0.8762
     & Distill-VQ & \textbf{0.3523}$^\dagger$ & \textbf{0.6212}$^\dagger$ & \textbf{0.8086} & \textbf{0.8612}$^\dagger$ & \textbf{0.5781}$^\dagger$ & \textbf{0.7654}$^\dagger$ & \textbf{0.8543}$^\dagger$ & \textbf{0.8762}$^\dagger$  \\
    \ChangeRT{1pt}
    \end{tabular}
    % \vspace{-5pt}
    \caption{\small Overall Performances on MS MARCO Passage Retrieval and Natural Questions, with embeddings from AR2-G and CoCondenser.  $^\dagger$ indicates the improvement over the strongest baseline is statistically significant on a paired t-test (p < 0.05).}
    \vspace{-10pt}
    \label{tab:2}
\end{table*}

\section{Experiments}
\subsection{Experiment Settings}

\subsubsection{Datasets}
Our experimental studies are based on the two popular benchmarks on document retrieval. The first one is the \textbf{MS MARCO (passage retrieval)} \cite{nguyen2016ms}\footnote{Following the existing practice, MS MARCO is tested on its released validation set.}. It is a widely used benchmark on web search and embedding-based retrieval, where the targeted answers from MS MARCO corpus need to be retrieved for queries from Bing search. The second one is the \textbf{Natural Questions (NQ)} \cite{kwiatkowski2019natural}. The queries are real-world questions collected from Google Search; each of the queries is paired with an answer span and ground-truth passages from the Wikipedia pages. The detailed specifications of the datasets are shown as Table \ref{tab:1}. For both datasets, we expect the ground-truth answer to each query can be retrieved from the entire corpus. Therefore, the experiment performances will be measured by the recall rate for the top-K retrieval result. Besides, we'll also report {MRR@10} for more comprehensive evaluation. 

\subsubsection{Baseline methods}
We consider the following groups of baselines in experiments (corresponding to the discussions in Section \ref{sec:ref}). Firstly, the typical vector quantization methods: \textbf{IVFPQ} \cite{jegou2011searching}, \textbf{IVFOPQ} \cite{ge2013optimized}, and \textbf{ScaNN} \cite{guo2020accelerating}, where the vector quantization is learned for reconstruction loss minimization with fixed document embeddings. Secondly, the latest work on the joint learning of embedding models and vector quantization for reconstruction loss minimization:  \textbf{Poeem} \cite{zhang2021joint}. Thirdly, the latest work on the joint learning of embedding models and vector quantization for retrieval loss minimization: \textbf{JPQ} \cite{zhan2021jointly} and \textbf{RepCONC} \cite{zhan2021learning}. Note that JPQ and RepCONC utilize the original IVF provided by FAISS rather than learn it as the PQ component. 
We further introduce a variant of \textbf{MoPQ}\cite{xiao2021matching} for ablation study, which optimizes the IVF and PQ both by contrastive learning. 

\subsubsection{Implementation details} Distill-VQ (our approach) leverages BERT-based backbone \cite{devlin2018bert} for text encoding. In our experiments, we work with the released well-trained encoders from the latest works: \textbf{AR2-G} \cite{zhang2021adversarial} and \textbf{CoCondenser} \cite{gao2021unsupervised}, which are the most accurate models on MS MARCO and NQ benchmarks. For the sake of fair comparisons, all the baseline methods will use \textbf{the same document embeddings} as Distill-VQ (which will be more accurate than their original implementations). For PQ, OPQ, ScaNN: the quantization will be directly applied to document embeddings from AG2-G and CoCondenser; for Poeem: the query encoder and the quantization module will be jointly learned for reconstruction loss minimization; for JPQ, RepCONC, MoPQ: the query encoder and the quantization module will be jointly learned for retrieval loss minimization. We wrap up the learned quantization results with the IVFPQ index implemented by \textbf{FAISS} \cite{johnson2019billion}, which enables document retrieval to be efficiently performed. The default configuration of index is specified as follows. For IVF: a total of 10,000 posting lists are deployed, of which the Top-100 posting lists (i.e., the Top 1$\%$) will be visited for ANN search; for PQ, a total of 64 codebooks are deployed, each of which uses 8-bit codewords (i.e., 256 codewords per codebook). Aside from the default configuration, which we will also explore more settings of IVF and PQ in our experiments.  

The experiments are based on clusters of 8* NVIDIA-A100-40GB GPUs and 2* AMD EPYC 7V12 64-Core CPUs. The algorithms are implemented with python 3.6 and PyTorch 1.8.0. 
We optimize the parameters with the AdamW optimizer. 
The learning rates for query encoder, IVF centroids and PQ codebooks are 5e-6, 1e-3 and 1e-4 respectively.
More training details can be found in our released code: 
https://github.com/staoxiao/LibVQ.
% The training time of Distill-VQ is 1.1 hours on MSMARCO dataset and 0.3 hours on NQ dataset.

% Our source code, models and datasets will all be open to public. 

\subsubsection{Problems to be explored}
Our experiments focus on three problems. First and foremost, Distill-VQ's impact on retrieval quality against the existing VQ methods (Section \ref{sec:exp-overall}). Secondly, analysis of knowledge distillation: the impact from different similarity functions and different sampling strategies for candidate documents (Section \ref{sec:exp-kd}). Thirdly, exploration of Distill-VQ's impact on each individual component: IVF and PQ, respectively, with variant configurations of both components (Section \ref{sec:exp-ivf}, \ref{sec:exp-pq}). 

\begin{table*}[h]
    \centering
    \footnotesize
    \begin{tabular}{p{2.6cm} | C{1.2cm} C{1.2cm} C{1.2cm} C{1.2cm} | C{1.2cm} C{1.2cm} C{1.2cm} C{1.2cm} }
    \ChangeRT{1pt}  & 
    \multicolumn{4}{c|}{\textbf{MS MARCO Passage (Dev)}} & \multicolumn{4}{c}{\textbf{Natural Question (Test)}}   \\
    \cmidrule(lr){1-1}
    \cmidrule(lr){2-5}
    \cmidrule(lr){6-9}
     \textbf{Method} & 
     \textbf{MRR@10} & \textbf{R@10} & \textbf{R@50} & \textbf{R@100} & \textbf{MRR@10} & \textbf{R@10} & \textbf{R@50} & \textbf{R@100}  \\
     \hline
     MSE & 0.3518 & 0.6158 & 0.7981 & 0.8457 & 0.5725 & 0.7545 & 0.8313 & 0.8520 \\
     Margin-MSE & 0.3592 & 0.6207 & 0.8107 & 0.8595 & 0.5869 & 0.7573 & 0.8324 & 0.8504 \\
     RankNet & 0.3519 & 0.6156 & 0.8044 & 0.8560 & 0.6052 & 0.7844 & 0.8623 & 0.8764 \\
     KL-Div & \textbf{0.3633} & 0.6239 & 0.8153 & 0.8612 & 0.5846 & 0.7576 & 0.8315 & 0.8506 \\
     ListNet & 0.3607 & \textbf{0.6276} & \textbf{0.8221} & \textbf{0.8719} & \textbf{0.6235} & \textbf{0.7927} & \textbf{0.8627} & \textbf{0.8783} \\
     \hline
     GT+IB & 0.3577 & 0.6240 & 0.8194 & 0.8710 & 0.6224 & 0.7910 & 0.8600 & 0.8703 \\
     {GT+IB+Top-200$^1$} & 0.3607 & 0.6276	& 0.8221 & 0.8719 & 0.6235 & 0.7927	 & 0.8627 & 0.8783 \\
     IB (*) & 0.3543 & 0.6223 & 0.8115 & 0.8608 & 0.5992 & 0.7731 & 0.8457 & 0.8642 \\
     Top-3$^1$+Top-200$^1$ (*) & 0.3595	& 0.6215 & 0.8084 & 0.8541	& 0.5941 & 0.7603 & 0.8360 & 0.8545 \\
     IB+Top-200$^2$ (*) & 0.3598 & 0.6237 & 0.8126 & 0.8616 & 0.6186 & 0.7883 & 0.8479 & 0.8662 \\
     IB+Top-3$^1$+Top-200$^1$ (*) & 0.3603 & 0.6265 & \textbf{0.8221} & 0.8727 & 0.6228 & 0.7915 & 0.8624 & 0.8786 \\
     IB+Top-10 (*) & 0.3614 & 0.6285 & 0.8201 & 0.8698 & 0.6130 & 0.7853 & 0.8581 & 0.8753 \\
     IB+Top-100 (*) & 0.3655 & 0.\textbf{6312} & 0.8210 & 0.8690 & 0.6271 & 0.7933 & 0.8637 & 0.8800 \\
     IB+Top-200 (*) & \textbf{0.3656} & 0.6298 & 0.8205 & \textbf{0.8871} & \textbf{0.6295} & \textbf{0.7969} & \textbf{0.8675} & \textbf{0.8806} \\
    %  Top-10 as GT (Contrast) & 0.2526 & 0.4440 & 0.5943 & 0.6535 & 0.4188 & 0.6119 & 0.7603 & 0.8005 \\
    %  Top-100 as GT (Contrast) & 0.3570 & 0.6204 & 0.8127 & 0.8600 & 0.5870 & 0.7723 & 0.8584 & 0.8734 \\
    %  Top-200 as GT (Contrast) & 0.3569 & 0.6225 & 0.8103 & 0.8600 & 0.5856 & 0.7706 & 0.8565 & 0.8734 \\
    \ChangeRT{1pt}
    \end{tabular}
    % \vspace{-5pt}
    \caption{\small Knowledge distillation analysis. Upper: impact from similarity functions. Lower: impact from sampled documents (GT: ground-truth; IB: in-batch sampling; Top-K$^1$/Top-K$^2$: sampling 1/2 document from Top-K; Top-K: using all Top-K documents; (*): w.o. ground-truth.)}
    % Contrast: performing contrast learning by using the Top-K documents as ground-truth.
    \vspace{-20pt}
    \label{tab:6}
\end{table*}

\subsection{Experiment Analysis} 

%% Distill-VQ's general advantages: 
%% sota, *48 compression, *100 speedup
%% Learnable's advantages over fixed
%% Retrieval's advantage over reconstruct 
%% why JPQ and RepCONC inferior 

\subsubsection{Overall performance}\label{sec:exp-overall}
The overall performances of different methods are evaluated under the default settings (IVF: search Top-100 out of 10,000 posting lists; PQ: 64 codebooks, each codebook is 8-bit). All the methods utilize the same document embeddings from AR2-G and CoCondener for fair comparisons. Distill-VQ's general effectiveness can be verified given its huge advantages over the baselines (as Table \ref{tab:2}). For MS MARCO: Distill-VQ achieves 0.3607 and 0.3523 on MRR@10 with AR2-G and CoCondenser, outperforming the strongest baselines by +1.36\% and +1.53\%, respectively. For NQ: Distill-VQ achieves 0.6235 and 0.5781 on MRR@10 with both embeddings, leading to +4.0\% and +0.92\% over the strongest baselines. The above performances are also the SOTA results of their kinds as compared with the reported results in the latest works \cite{zhan2021jointly,zhan2021learning,xiao2022progressively}. Note that our evaluation is performed on top of IVF with merely 1\% of the posting lists searched, which means there is a $100\times$ acceleration over the previous works. 

Besides, it is also interesting to find that although the joint learning baselines (Poeem, JPQ, RepCONC, MoPQ) are still generally better than PQ, OPQ and ScaNN, their advantages over IVFOPQ are not as significant as the observations in previous works \cite{zhan2021jointly,zhan2021learning,xiao2022progressively}. Meanwhile, the performances of Poeem (minimizing the reconstruction loss) and other joint learning methods (minimizing the retrieval loss) are also close to each other in many cases. Much of these observations is probably due to the enhancement of document embeddings, as AR2-G and CoCondenser are more accurate than conventional models, like DPR \cite{karpukhin2020dense}, ANCE \cite{xiong2021approximate}. However, the joint learning methods for retrieval loss minimization may regain notable advantages when working with larger acceleration and compression ratios, as to be discussed in Section \ref{sec:exp-ivf} and \ref{sec:exp-pq}.

\begin{table*}[h]
    \centering
    \footnotesize
    \begin{tabular}{p{1.6cm} | p{2.1cm} | C{1.2cm} C{1.2cm} C{1.2cm} C{1.2cm} C{1.2cm} C{1.2cm} C{1.2cm} }
    \ChangeRT{1pt}  & & 
    \multicolumn{7}{c}{\textbf{Recall@100 with different number of posting lists to search}}   \\
    \cmidrule(lr){1-1}
    \cmidrule(lr){2-2}
    \cmidrule(lr){3-9}
     \textbf{Dataset} & \textbf{Method} & \textbf{Top-1} & 
     \textbf{Top-10} & \textbf{Top-30} & \textbf{Top-50} & \textbf{Top-100} & \textbf{Top-300} & \textbf{Top-500} 
     \\
     \hline
     \multirow{4}{*}{MS MARCO}
     & IVF + OPQ & 0.4378 & 0.5299 & 0.5744 & 0.8180 & 0.8385 & 0.8658 & {0.8737} \\
     & IVF (Rctr) + OPQ & 0.4550 & 0.7394 & 0.7988 & 0.8158 & 0.8366 & 0.8610 & 0.8687\\
     & IVF (Rtrv) + OPQ & 0.5018 & 0.7673 & 0.8147 & 0.8303 & 0.8493 & 0.8630 & 0.8690 \\
     & IVF (Distill) + OPQ & \textbf{0.5203} & \textbf{0.7807} & \textbf{0.8289} & \textbf{0.8442} & \textbf{0.8614} & \textbf{0.8789} & \textbf{0.8838} \\
     \hline
     \multirow{4}{*}{NQ} 
     & IVF + OPQ & 0.5168 & 0.7080 & 0.7731 & 0.7936 & 0.8279 & 0.8634 & 0.8759 \\
     & IVF (Rctr) + OPQ & 0.5792 & 0.7484 & 0.7986 & 0.8218 & 0.8443 & 0.8684 & 0.8778 \\
     & IVF (Rtrv) + OPQ & 0.6022 & 0.7750 & 0.8235 & 0.8401 & 0.8556 & 0.8711 & 0.8767 \\
     & IVF (Distill) + OPQ & \textbf{0.6875} & \textbf{0.8326} & \textbf{0.8612} & \textbf{0.8706} & \textbf{0.8736} & \textbf{0.8875} & \textbf{0.8903} \\ 
    \ChangeRT{1pt}
    \end{tabular}
    % \vspace{-5pt}
    \caption{\small Distill-VQ's impact on IVF with different number of posting lists to search. The top-K (increased from 1 to {500}, with acceleration ratio changing from 10,000$\times$ to {20$\times$}) out of the 10,000 deployed posting lists are searched for document retrieval.}
    \vspace{-15pt}
    \label{tab:4} 
\end{table*}

\begin{table*}[h]
    \centering
    \footnotesize
    \begin{tabular}{p{1.6cm} | p{2.0cm} | C{1.2cm} C{1.2cm} C{1.2cm} C{1.2cm} C{1.2cm} C{1.2cm} C{1.2cm} C{1.2cm} }
    \ChangeRT{1pt}  & & 
    \multicolumn{7}{c}{\textbf{Recall@100 with different number of deployed posting lists}}   \\
    \cmidrule(lr){1-1}
    \cmidrule(lr){2-2}
    \cmidrule(lr){3-9}
     \textbf{Dataset} & \textbf{Method}& 
     \textbf{100} & \textbf{500} & \textbf{1,000} & \textbf{5,000} & \textbf{10,000} & \textbf{50,000} & \textbf{100,000} \\
     \hline
     \multirow{4}{*}{MS MARCO} 
     & IVF + OPQ & 0.5424 & 0.7525 & 0.7828 & 0.8231 & 0.8385 & 0.8639 & 0.8728 \\
     & IVF (Rctr) + OPQ & 0.5390 & 0.7567 & 0.7908 & 0.8305 & 0.8366 & 0.8642 & 0.8735 \\
     & IVF (Rtrv) + OPQ & 0.5916 & 0.7757 & 0.8083 & 0.8476 & {0.8493} & 0.8724 & 0.8795 \\
     & IVF (Distill) + OPQ & \textbf{0.6123} & \textbf{0.7900} & \textbf{0.8200} & \textbf{0.8524} & \textbf{0.8614} & \textbf{0.8801} & \textbf{0.8850} \\
     \hline
     \multirow{4}{*}{NQ} 
     & IVF + OPQ & 0.5814 & 0.7437 & 0.7700 & 0.8113 & 0.8279 & 0.8506 & 0.8592 \\
     & IVF (Rctr) + OPQ & 0.5814 & 0.7581 & 0.7872 & 0.8337 & 0.8443 & 0.8604 & 0.8599 \\
     & IVF (Rtrv) + OPQ & 0.7202 & 0.8132 & 0.8318 & 0.8487 & 0.8556 & 0.8609 & 0.8573 \\
     & IVF (Distill) + OPQ & \textbf{0.7858} & \textbf{0.8445} & \textbf{0.8556} & \textbf{0.8725} & \textbf{0.8736} & \textbf{0.8819} & \textbf{0.8836} \\ 
    \ChangeRT{1pt}
    \end{tabular}
    % \vspace{-5pt}
    \caption{\small Distill-VQ's impact on IVF with different number of deployed posting lists. The number of deployment is increased from 100 to 100,000; 1\% of the posting lists will be searched (e.g., Top-100 for 10,000) so that the acceleration ratio is always $100\times$.}
    \vspace{-15pt}
    \label{tab:5}
\end{table*}

\subsubsection{Explorations of knowledge distillation}\label{sec:exp-kd} We further explore the similarity function and candidate documents' impact on knowledge distillation. The remaining experiments will be based on AR2-G unless otherwise specified, given the similar observations between the two embeddings. 

$\bullet$ \textbf{Similarity function's impact}
The evaluation results are shown in the upper part of Table \ref{tab:6}, where the five alternative functions discussed in Section \ref{sec:sim_fn}: MSE, Margin-MSE, RankNet, KL-divergence, and ListNet, are compared against each other. It can be observed that different similarities do lead to distinct distillation effects, where the overall performances from enforcing the ranking order invariance (KL-Div, ListNet, RankNet) are comparatively better than those enforcing the score invariance (MSE, Margin-MSE). As discussed, the ranking order invariance is a sufficient condition for preserving dense embeddings' retrieval performance. Besides, it is a largely relaxed requirement compared with the score invariance, which can be distilled more easily. The same reasons may also be applied to explain Margin-MSE's advantage over MSE. We may also observe that ListNet remains competitive on both of the datasets; in contrast, KL-Div and RankNet only achieve competitive performance on one of the datasets. For RankNet, the unnormalized scores ($\mathbf{t}_{q,d}$) from teachers are directly utilized; for KL-Div, the students' distribution ($\Phi^s_{q,*}$) is implicitly regularized to be uniform (whereas the teachers' distribution can be skewed in reality). Both issues may become adversarial factors for optimization under specific conditions of the teachers.

\begin{table*}[h]
    \centering
    \footnotesize
    \begin{tabular}{p{1.4cm} | p{2.1cm} | C{1.0cm} C{1.0cm} | C{1.0cm} C{1.0cm} | C{1.0cm} C{1.0cm} | C{1.0cm} C{1.0cm} | C{1.0cm} C{1.0cm} }
    \ChangeRT{1pt}  & & \multicolumn{2}{c|}{2$\times$8} & \multicolumn{2}{c|}{4$\times$8} & \multicolumn{2}{c|}{8$\times$8} & \multicolumn{2}{c|}{16$\times$8} & \multicolumn{2}{c}{32$\times$8} \\
    \cmidrule(lr){1-1}
    \cmidrule(lr){2-2}
    \cmidrule(lr){3-4}
    \cmidrule(lr){5-6}
    \cmidrule(lr){7-8}
    \cmidrule(lr){9-10}
    \cmidrule(lr){11-12}
     \textbf{Dataset} & \textbf{Method} & 
     MRR@10 & R@100 & MRR@10 & R@100 & MRR@10 & R@100 & MRR@10 & R@100 & MRR@10 & R@100 \\
     \hline
     \multirow{7}{*}{MS MARCO} 
     & IVF + PQ & 0.0066 & 0.0903 & 0.0051 & 0.0825 & 0.0094 & 0.1059 & 0.0489 & 0.2896 & 0.1431 & 0.5546 \\
     & IVF + OPQ & 0.0123 & 0.1355 & 0.0444 & 0.3027 & 0.0772 & 0.4201 & 0.2599 & 0.7558 & 0.3220 & 0.8183 \\
     & IVF + ScaNN & 0.0187 & 0.1559 & 0.0336 & 0.3220 & 0.0497 & 0.3778 & 0.1147 & 0.5273 & 0.1883 & 0.6553 \\
     & IVF + PQ (Poeem) & 0.0122 & 0.1268 & 0.0459 & 0.2986 & 0.0934 & 0.4367 & 0.2760 & 0.7574 & 0.3253 & 0.8213 \\
     & IVF + PQ (JPQ) & 0.0146 & 0.1350 & 0.0785 & 0.4412 & 0.1611 & 0.6273 & 0.2897 & 0.8118 & 0.3308 & 0.8351 \\
     & IVF + PQ (RepCONC) & 0.0195 & 0.1760 & 0.0760 & 0.4371 & 0.1400 & 0.5789 & 0.2817 & 0.8048 & 0.3284 & 0.8372 \\
     & IVF + PQ (MoPQ) & 0.0231 & 0.2120 & 0.0923 & 0.4855 & 0.1621 & 0.6222 & 0.2935 & 0.8089 & 0.3309 & 0.8406 \\
     & IVF + PQ (Distill) & \textbf{0.0241} & \textbf{0.2406} & \textbf{0.1022} & \textbf{0.5274} & \textbf{0.1773} & \textbf{0.6476} & \textbf{0.3048} & \textbf{0.8106} & \textbf{0.3399} & \textbf{0.8410} \\
     \hline
     \multirow{7}{*}{NQ} 
     & IVF + PQ & 0.0450 & 0.3479 & 0.0493 & 0.3523 & 0.0547 & 0.3679 & 0.0762 & 0.4457 & 0.1404 & 0.5673 \\
     & IVF + OPQ & 0.0780 & 0.4254 & 0.1336 & 0.4986 & 0.1572 & 0.5772 & 0.3517 & 0.7565 & 0.4819 & 0.8127 \\
     & IVF + ScaNN & 0.0828 & 0.4306 & 0.1521 & 0.5473 & 0.1610 & 0.5925 & 0.2014 & 0.6470 & 0.2923 & 0.7293 \\
     & IVF + PQ (Poeem) & 0.0783 & 0.4224 & 0.1375 & 0.5198 & 0.1602 & 0.5912 & 0.3669 & 0.7686 & 0.4901 & 0.8244 \\
     & IVF + PQ (JPQ) & 0.0675 & 0.4138 & 0.1587 & 0.5529 & 0.1937 & 0.6508 & 0.4080 & 0.8132 & 0.5040 & 0.8324 \\
     & IVF + PQ (RepCONC) & 0.0847 & 0.4299 & 0.1452 & 0.5307 & 0.1853 & 0.6227 & 0.4133 & 0.8121 & 0.4988 & 0.8376 \\
     & IVF + PQ (MoPQ) & 0.0917 & 0.4592 & 0.1642 & 0.5756 & 0.2163 & 0.6716 & 0.4613 & 0.8229 & 0.5220 & 0.8396 \\
     & IVF + PQ (Distill) & \textbf{0.0945} & \textbf{0.4889} & \textbf{0.1802} & \textbf{0.6254} & \textbf{0.2661} & \textbf{0.7047} & \textbf{0.4776} & \textbf{0.8271} & \textbf{0.5401} & \textbf{0.8418} \\ 
    \ChangeRT{1pt}
    \end{tabular}
    % \vspace{-5pt}
    \caption{\small Distill-VQ' Impact on PQ, with the size of codebooks expanded from 2$\times$8 to 32$\times$8 (\#Codebooks $\times$ Bit\_Per\_Codebook). All methods in comparison use the same IVF generated by FAISS (search Top-100 out of 10,000 posting lists for $100\times$ acceleration).} 
    \vspace{-10pt}
    \label{tab:3}
\end{table*}

\begin{figure*}[t]
\centering
\includegraphics[width=0.92\textwidth]{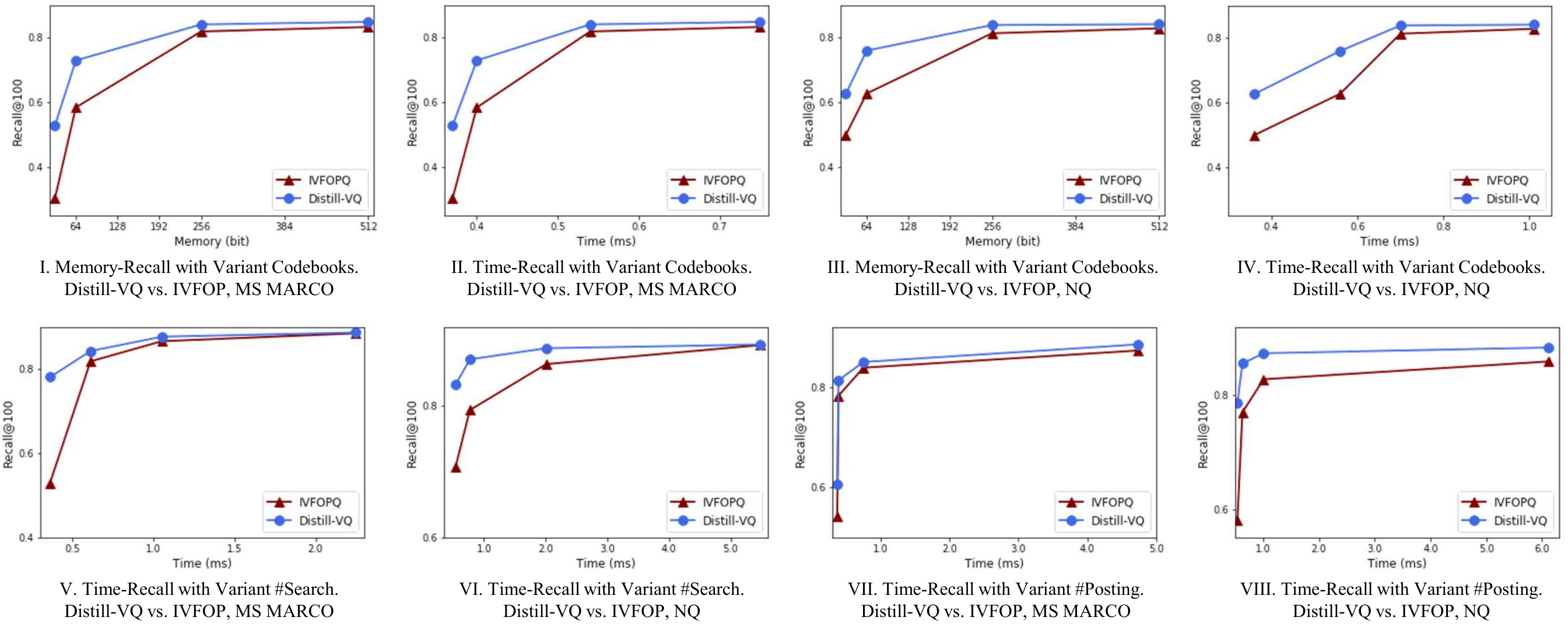}
\vspace{-10pt}
\caption{\small Efficiency and Retrieval Quality. I-IV: Recall@100 at different time/memory efficiency by changing the size of codebooks. V-VIII: Recall@100 at different time efficiency by changing 1) \#Search: search the Top-\#Search out of the 10,000 deployed posting lists, and 2) \#Posting: the number of deployed posting lists (with the Top-1\% searched).}
\vspace{-10pt}
\label{fig:3}
\end{figure*}

$\bullet$ \textbf{Candidate documents' impact} The analysis for candidate documents' impact are shown as the lower part of Table \ref{tab:6}. In this place, the candidate documents come from the combinations of the following sampling sources: 1) using the ground-truth document, labeled as GT; 2) using in-batch document sampling, labeled as IB; 3) randomly sampling one/two candidate from the Top-K documents, labeled as Top-K$^1$/Top-K$^2$ (e.g., Top-3$^1$ means sample one document from the Top-3, and Top-200$^2$ means sampling two documents from the Top-200); 4) using all Top-K documents, labeled as Top-K. We may have the following findings from the demonstrated results. 

Firstly, the labeled data is no longer a necessity in knowledge distillation, and highly competitive retrieval performance can be achieved by distilling knowledge from the unlabeled data (marked with ``*''). We may observe that the combination of in-batch sampling (IB) and Top-K documents outperforms the other sampling strategies in comparison, especially those simply working with ground-truth and in-batch. Besides, it is also interesting to find that the retrieval quality tends to benefit more from using an increased amount of Top-K documents: using the entire Top-200 is much more effective than simply using a few samples from them (IB+Top-200 against IB+Top-200$^2$), and using Top-100 and Top-200 outperform the one simply using Top-10 (IB+Top-100/200 against IB+Top-10).

% highly competitive retrieval performance can be derived by distilling knowledge from unlabeled data. In particular, the combination of in-batch sampling (IB) and Top-K documents outperforms all the sampling strategies in comparison. It is also interesting to find that the retrieval quality tends to benefit more from using an increased amount of Top-K documents: using the entire Top-200 is much more effective than simply using a few samples from them (comparing IB+Top-200 against IB+Top-200$^2$), and using Top-100 and Top-200 outperform the one simply using Top-10 (comparing IB+Top-100/200 against IB+Top-10).  

% Secondly, the labeled data is no longer a necessity in knowledge distillation, knowing that most of the unlabeled knowledge distillation (marked by ``*'') may give rise to high

% as IB+Top-3$^1$+Top-200$^1$ (without label) gives rise to on-par performance as GT+IB+Top-200$^1$ (using labeled data). Besides, the Top-K documents also serve as notable supplement to the ground-truth: by leveraging one more document sampled from Top-200, GT+IB+Top-200' is able to outperform GT+IB on both datasets. 

Secondly, although the top-ranked documents are important, the lower-ranked documents from in-batch sampling are still necessary for knowledge distillation. Comparing Top-3$^1$+Top-200$^1$ with IB+Top-3$^1$+Top-200$^1$: the retrieval quality drops a lot on both datasets when in-batch sampling is disabled. However, it is also undesirable to simply work with the in-batch documents due to the lack of top-ranked documents, as the performance from IB alone is much lower than other methods. 

% Finally, the exploitation of the Top-K documents is a unique merit of knowledge distillation. Although it is trivial to make a similar adaptation for the contrastive learning, where the Top-K documents are utilized as the ground-truth, the resulted performances (``Top-K as GT (Contrast)'') are quite limited compared with their counterparts from knowledge distillation (IB+Top-K). 

\vspace{-2pt}
\subsubsection{Distill-VQ's impact on IVF}\label{sec:exp-ivf}
Detailed exploration of Distill-VQ's impact on IVF is shown in Table \ref{tab:4} and \ref{tab:5}. To focus on the impact solely from IVF, all methods in comparison utilize the same product quantization: OPQ, as implemented by FAISS. A total of four methods are compared: 1) IVF: the original IVF implemented by FAISS, 2) IVF (Rctr): the IVF learned for reconstruction loss minimization (Poeem), 3) IVF (Rtrv): the IVF learned for retrieval loss minimization (MoPQ), and 4) IVF (Distill): the IVF learned from knowledge distillation (Distill-VQ). 

For evaluations in Table \ref{tab:4}, a total of 10,000 posting lists are deployed in IVF. The Top-K posting lists are searched for document retrieval, where a smaller K indicates a larger acceleration ratio, e.g., Top-10 indicates 1,000$\times$ speedup up (10,000/10). According to the experiment results, the retrieval quality can be continuously increased for all the methods with the growth of K; during this process, IVF (Distill) maintains notable advantages over other baselines. Besides, IVF (Distill) achieves more significant improvements when larger acceleration ratios are adopted (i.e., smaller K): when K equals to 1, IVF (Distill) outperforms the baselines by +2$\sim$6\% on MS MARCO, and +8$\sim$17\% on NQ. The advantages gradually diminish to +1\% on both datasets when K increases to 500. 

It is also interesting to see that both IVF (Rtrv) and IVF (Rctr) are no better than the original IVF when a large number of posting lists can be searched (e.g., K=500). This is probably because the jointly learning methods target on optimizing the Top-1 posting list of each query, whereas caring little of the retrieval quality from the Top-K posting lists.

% For MS MARCO: IVF (Distill) consistently outperforms the strongest baseline by +1\%$\sim$+2\%. For NQ: IVF (Distill) achieves significant improvements when K is small; e.g., Recall@100 can be improved by 8.53\% when K is 1. Although the improvements diminish when K becomes larger, it still maintains around 2\% advantages over the strongest baselines. 

Secondly, the number of deployed posting lists is increased from 100 to 100,000 in Table \ref{tab:5}. The document retrieval will always be made by searching the top 1\% posting lists (e.g., the Top-1 will be searched for 100, and the Top-50 will be searched for 5,000), thus keeping a fixed acceleration ratio of 100$\times$. For both MS MARCO and NQ, IVF (Distill) achieves notable improvements over the strongest baselines in every setting. Besides, it is interesting to observe that although the acceleration ratio is fixed, the performances are gradually improved when more posting lists are deployed. Since an observation is consistent with the previous finding \cite{babenko2014inverted,baranchuk2018revisiting} that fine-grained inverted indexes are beneficial to retrieval quality. The search for the Top-K indexes can be expensive when a large number of posting lists are deployed; in \cite{baranchuk2018revisiting}, the problem is mitigated by establishing another ANN index (e.g., HNSW) for the posting lists.

%% Settings
%% Distill-VQ advantages in large compression ratio
%% Performance come close in lower compression ratio

%% Reconstruct < Retrieval 
\vspace{-6pt}
\subsubsection{Distill-VQ's impact on PQ}\label{sec:exp-pq}
Further exploration of Distill-VQ's impact on PQ is made as Table \ref{tab:3}. To focus on the impact solely from PQ, all methods in comparison utilize the same IVF generated by FAISS. The number of the codebooks is increased from 2 to 32: a smaller scale of codebooks means a larger compression ratio. With the growth of codebooks scale, the retrieval quality can be consistently improved for all the methods. When small-scale codebooks are used (meaning that the compression ratio is large), Distill-VQ is able to outperform all the baselines with notable advantages in both Recall and MRR. Besides, we may also compare different baselines more clearly when small-scale codebooks are used. Particularly, the jointly learning methods (Poeem, JPQ, RepCONC, MoPQ) achieve more significant and consistent advantages against the conventional methods (PQ, OPQ); and the methods on retrieval loss minimization (JPQ, RepCONC, MoPQ) are superior than the one minimizing the reconstruction loss (Poeem). Finally, Distill-VQ's relative advantages gradually diminish; when codebooks become sufficiently large, some of the baselines (JPQ, RepCONC, MoPQ) are able to get close to Distill-VQ in terms of Recall. This is because larger codebooks are more expressive than the smaller ones, which makes it easier to preserve the dense embeddings' retrieval performance. However, Distill-VQ still maintains notable advantages in MRR@10 (more discriminative than Recall@100), indicating that the corresponding VQ is still much better than other baselines.

\subsubsection{Efficiency and retrieval quality}\label{sec:exp-efficiency}
We further analyze Distill-VQ's retrieval quality at different working efficiency (Figure \ref{fig:3}). The memory efficiency is measured by the bit-rate of each document; e.g., 64-bit is corresponding to the 8$\times$8 codebooks, which leads to a $192\times$ compression of the original dim-768 float-32 document embeddings. The time efficiency is measured by the latency of getting the Top-100 documents for each query. The evaluation is performed on top of FAISS, where the vector quantization results from Distill-VQ are loaded into the IVFPQ index in FAISS. The original IVFOPQ from FAISS is introduced for comparison. 

In Figure \ref{fig:3} I-IV, Recall@100 is measured at different time and memory cost by changing the codebook size (correlation between recall and codebook size is discussed in Table \ref{tab:3}). In Figure \ref{fig:3} V-VIII, Recall@100 is measured at different time cost by changing \#Search: the amount of posting lists to search (10,000 deployed in total), and \#Posting: the number of posting lists to deploy, with the Top 1\% to be searched (correlations between recall and both factors are discussed in Table \ref{tab:4}, \ref{tab:5}). It can be observed that Distill-VQ outperforms IVFOPQ in all of these settings. Besides, the advantages become much more significant at small bit-rates and time latency, reflecting Distill-VQ's importance in massive-scale scenarios where high compression and acceleration ratios are needed. 

\vspace{-1pt}
\subsubsection{Summary of findings} The major findings of the experiments are summarized into the following points.
\begin{itemize}
    \item Disitll-VQ notably outperforms the existing vector quantization methods on MS MARCO and Natural Questions. 
    \item Both IVF and PQ modules may benefit from Disitll-VQ; the resulted advantages are more significant when larger compression and acceleration ratios are adopted. 
    \item The knowledge distillation can be effectively performed based on ListNet (enforcing ranking order invariance) and documents sampled from Top-K+In-Batch (w.o. using labeled data). 
    \item Distill-VQ results in better efficiency-recall trade-off as compared with the original IVFPQ in FAISS library. 
\end{itemize}

% \textbf{1.} Disitll-VQ notably outperforms the existing vector quantization methods on MS MARCO and Natural Questions. \textbf{2.} Both IVF and PQ modules may benefit from Disitll-VQ; the resulted advantages are more significant when larger compression and acceleration ratios are adopted. \textbf{3.} The knowledge distillation can be effectively performed based on ListNet (enforcing ranking order invariance) and documents sampled from Top-K+In-Batch (w.o. using labeled data). \textbf{4.} Distill-VQ results in better efficiency-recall trade-off as compared with the original IVFPQ in FAISS library. 

\section{Conclusion}
% We presented a novel framework Distill-VQ, which jointly learns IVF and PQ for the optimization of their retrieval performances. In Distill-VQ, the VQ modules are learned by distilling knowledge from the well-trained dense embeddings, which effectively exploits the massive-scale of unlabeled data for the improvement of VQ's performance. Comprehensive evaluations are performed on MS MARCO and Natural Questions benchmarks, whose results verify Distill-VQ's notable advantages over the SOTA baselines. 

We present a novel framework Distill-VQ, which jointly learns IVF and PQ for the optimization of their retrieval performances. In our framework, the VQ modules are learned by distilling knowledge from the well-trained dense embeddings, which enables massive-scale of unlabeled data to be exploited for the improvement of VQ's performance. Our empirical explorations indicate that Distill-VQ can be effectively conducted by enforcing teachers-students ranking order invariance (ListNet) over documents sampled from unlabeled data (Top-K+In-Batch). 
% The evaluations on MS MARCO and NQ verify Distill-VQ effectiveness in comparison with the SOTA VQ baselines. 
Knowing that the proposed method may conveniently work with those off-the-shelf embeddings and easily be integrated with FAISS, we expect it to serve as a generic tool to facilitate high-quality and efficient document retrieval.

\begin{acks}
This work is supported by the National Natural Science Foundation of China (Nos. U1936104, 62192784, 62022077 and 61976198) and CCF-Baidu Open Fund.
\end{acks}

\nocite{xiao2022progressively,lian2020lightrec,wu2021linear,jiang2021xlightfm}

\bibliographystyle{ACM-Reference-Format}
\bibliography{reference}

\end{document}